\documentclass[12pt]{iopart}

\usepackage{hyperref}
\hypersetup{
colorlinks = true,
allcolors = {blue},
pdfborder={0 0 0}, 
bookmarksnumbered=true,
pdftitle={Diffraction imaging of light induced dynamics in xenon-doped helium nanodroplets},
pdfauthor={Bruno Langbehn}
}

\usepackage[backend=biber,style=phys,biblabel=brackets,pageranges=false,articletitle=true,maxnames=10]{biblatex}
\DefineBibliographyStrings{english}{%
    andothers = {\etal}
}

\usepackage{graphicx}

\usepackage[usenames,dvipsnames]{xcolor}

\usepackage{mathabx}

\usepackage{textcomp}

\begin{document}

\title{Diffraction imaging of light induced dynamics in xenon-doped helium nanodroplets}

\author{B Langbehn$^1$, Y Ovcharenko$^{1,2}$, A Clark$^3$, M Coreno$^4$, R Cucini$^5$, A Demidovich$^6$, M Drabbels$^3$, P Finetti$^6$, M Di Fraia$^{6,4}$, L Giannessi$^6$, C Grazioli$^5$, D Iablonskyi$^7$, A C LaForge$^8$, T Nishiyama$^9$, V Oliver \'Alvarez de Lara$^3$, C Peltz$^{10}$, P Piseri$^{11}$, O Plekan$^6$, K Sander$^{10}$, K Ueda$^7$, T Fennel$^{10,12}$, K C Prince$^{6,13}$, F Stienkemeier$^8$, C Callegari$^{6,4}$, T M\"oller$^1$ and D Rupp$^{1,12,14}$}
\address{$^1$ Technische Universit\"at Berlin, Institut f\"ur Optik und Atomare Physik, 10623 Berlin, Germany}
\address{$^2$ European XFEL GmbH, 22607 Hamburg, Germany}
\address{$^3$ Laboratory of Molecular Nanodynamics, Ecole Polytechnique F\'ed\'erale de Lausanne (EPFL), 1015 Lausanne, Switzerland}
\address{$^4$ ISM-CNR, Istituto di Struttura della Materia, LD2 Unit, 34149 Trieste, Italy}
\address{$^5$ IOM-CNR, Istituto Officina dei Materiali, 34149 Trieste, Italy}
\address{$^6$ Elettra--Sincrotrone Trieste, S.C.p.A., 34149 Trieste, Italy}
\address{$^7$ Institute of Multidisciplinary Research for Advanced Materials, Tohoku University, Sendai 980-8577, Japan}
\address{$^8$ Physikalisches Institut, Universit\"at Freiburg, 79104 Freiburg, Germany}
\address{$^9$ Division of Physics and Astronomy, Graduate School of Science, Kyoto University, Kyoto 606-8502, Japan}
\address{$^{10}$ Institut für Physik, Universität Rostock, 18051 Rostock, Germany}
\address{$^{11}$ CIMAINA and Dipartimento di Fisica, Universit\`a degli Studi di Milano, 20133 Milano, Italy}
\address{$^{12}$ Max-Born-Institut f\"ur Nichtlineare Optik und Kurzzeitspektroskopie, 12489 Berlin, Germany}
\address{$^{13}$ Department of Chemistry and Biotechnology, Swinburne University of Technology, Victoria 3122, Australia}
\address{$^{14}$ Laboratorium f\"ur Festk\"orperphysik, ETH Z\"urich, 8093 Z\"urich, Switzerland}

\eads{\mailto{bruno.langbehn@physik.tu-berlin.de}, \mailto{thomas.moeller@physik.tu-berlin.de}, \mailto{daniela.rupp@phys.ethz.ch}}

\begin{abstract}

We have explored the light induced dynamics in superfluid helium nanodroplets with wide-angle scattering in a pump-probe measurement scheme. The droplets are doped with xenon atoms to facilitate the ignition of a nanoplasma through irradiation with near-infrared laser pulses. After a variable time delay of up to 800 ps, we image the subsequent dynamics using intense extreme ultraviolet pulses from the FERMI free-electron laser. The recorded scattering images exhibit complex intensity fluctuations that are categorized based on their characteristic features. 
Systematic simulations of wide-angle diffraction patterns are performed, which can qualitatively explain the observed features by employing model shapes with both randomly distributed as well as structured, symmetric distortions. This points to a connection between the dynamics and the positions of the dopants in the droplets. In particular, the structured fluctuations might be governed by an underlying array of quantized vortices in the superfluid droplet as has been observed in previous small-angle diffraction experiments. Our results provide a basis for further investigations of dopant-droplet interactions and associated heating mechanisms.
\end{abstract}


\section{Introduction}

The interaction of intense laser pulses with matter plays an important role in many fields, from modern high-tech applications such as laser machining \cite{Meijer2004} and microsurgery \cite{Quinto-Su2007}, extreme ultraviolet (XUV) lithography~\cite{Fomenkov2017}, or propulsion of small satellites~\cite{Phipps2010,Gurin2019} to fundamental and applied science. For example, in structure analysis of biological specimen using x-ray diffraction, radiation damage poses a substantial problem, thus fueling a considerable research interest in the processes accompanying and following the irradiation with intense x-ray bursts \cite{Howells2009}. 

In particular, atomic clusters can serve as isolated model systems of simple geometrical structure. Time-resolved studies, where an initial \emph{pump} light pulse excites the system and a subsequent \emph{probe} light pulse measures the state of the system after a defined temporal delay, enable observation of the light induced dynamics. In the extreme case, the free cluster will eventually fully disintegrate, as the energy deposited by the pump pulse cannot dissipate otherwise.
A detailed investigation of the fragments and the dynamics can give an insight into the processes and their inherent time scale, which, in comparison with theoretical models of the interaction, enable development of a fundamental understanding \cite{Bostedt2010,Arbeiter2011,Gorkhover2012,Rupp2016,Kumagai2018,Rupp2020}.
A basic picture divides the interaction of an intense light pulse with an atomic cluster into three steps~\cite{Saalmann2006,Fennel2010}: (i) the ionization of cluster atoms and emission of electrons until the increasing positive charge prevents further electrons from leaving the cluster compound, (ii) the formation and evolution of a nanoplasma of ions and quasi-free electrons on a short  timescale, often only a few femtoseconds, and (iii) an expansion leading to complete destruction of the cluster, accompanied by relaxation and recombination processes.

Helium clusters, often referred to as helium nanodroplets, are model systems particularly suited to such studies due to their very simple electronic structure and smooth behavior: They remain liquid down to absolute zero forming mostly spherical shapes \cite{Gomez2014,Langbehn2018}, and they are transparent from the far-infrared to the vacuum ultraviolet because of the large ionization potential of helium ($24.6~\mathrm{eV}$). 
Using a near infrared (NIR) laser pulse, helium nanodroplets can be ionized only when the power density is sufficiently high that multiphoton ionization or field driven processes like tunneling ionization become significant ($\sim10^{15}~\mathrm{W}\mathrm{cm}^{-2}$)~\cite{Augst1989,Bacellar2022}.
Another option is to dope the droplets using atoms with a low ionization potential: In this case, ionization becomes possible already at lower laser intensity~\cite{Mikaberidze2009,Krishnan2011}, starting an avalanche-like process at the dopant positions~\cite{Schutte2016,Heidenreich2016}.
Depending on the interaction of the impurity with the helium solvent, the dopants remain at the surface of the droplet or get immersed, as is the case for, e.g., xenon atoms \cite{Toennies2004}.
When xenon-doped droplets are irradiated by an NIR pulse, theory predicts a very efficient complete ionization even with only a few xenon atoms embedded~\cite{Mikaberidze2009}. For the simple model of a compact dopant cluster in the center of the helium droplet, an anisotropic growth of the nanoplasma along the laser polarization axis has been predicted upon ignition~\cite{Mikaberidze2009}. On the other hand, dopant cluster formation has been observed at \emph{multiple} centers in larger droplets~\cite{Loginov2011}. Further, the superfluid nature of the helium nanodroplets leads to the occurrence of quantized vortices in large droplets~\cite{Gomez2014}, and the dopants agglomerate along the vortex lines~\cite{Tanyag2015,Jones2016}. Nanoplasma ignition can therefore be expected in large droplets at multiple sites, randomly distributed or structured by the vortex positions, which might lead to complex dynamics. 
In addition to nanoplasma dynamics governed by Coulombic forces between charged constituents, geometric changes beyond homogeneous Coulombic explosion or hydrodynamic expansion might have to be considered in helium nanodroplets. For example, the formation of voids around free electrons, so-called \emph{electron bubbles}, has been described in liquid helium~\cite{Rosenblit1995}.  
Further, the formation of gas bubbles has been observed on a macroscopic scale around copper nanoparticles in bulk liquid helium~\cite{Fernandez2015} and has also been reported on a microscopic scale around silver clusters in helium nanodroplets~\cite{Jones2019}, in both cases induced by resonant heating of the metal particles with an optical laser. 
Although superfluid helium is known for its very large heat conductivity, the flow of thermal energy from a solid particle to liquid helium is hindered by a discrepancy in velocity of sound, an effect called Kapitza resistance~\cite{Jones2019,Pollack1969}, that leads to a sudden evaporation of the surrounding helium, eventually forming a gas bubble.
A third process that could be taken into account is the ejection of dopant ions from a helium nanodroplet, leaving the droplet with several helium atoms attached to them~\cite{Bonhommeau2008,Halberstadt2020}.
These considerations show that the study of light induced dynamics in doped helium droplets may have particularly interesting aspects. So far, most theoretical work has been carried out for relatively small helium droplets ($N\leq10^6$ atoms per droplet).
While imaging experiments inherently address larger droplets \cite{Gomez2014,Langbehn2018}, their outcome has already stimulated theoretical studies, e.g., on the droplet shapes \cite{Ancilotto2015,Ancilotto2018}. It has been the close interplay between theory and experiment that established a more profound knowledge of the underlying physics of rotating superfluid droplets \cite{OConnell2020}. 
Therefore, it would be particularly interesting to experimentally visualize the droplet dynamics after laser excitation, giving a basis for comparison with theoretical models. Eventually, if droplet ignition at multiple sites or geometric changes in the droplets are observed, such experiments might lead to an extension of the models or even to a new approach to gain further insight into the processes, e.g., by employing molecular dynamics simulations. 

In this work, we report on a study of the dynamics triggered by intense NIR laser pulses in large, xenon-doped helium nanodroplets.
In this context, the advent of short-wavelength free-electron lasers (FELs) has opened up a new route to determine the structure of individual nanoparticles via coherent diffraction imaging (CDI) \cite{Neutze2000}, which is based on recording the light scattered off a particle. In general, such a \emph{diffraction pattern} reflects the scattering behavior of the particle, i.e., it depends on the distribution of atoms and their response to the wavelength of the incident light. Therefore, a change in scattering strength, e.g., because of a modification of the particle's density or its electron distribution, leads to a change of the intensity distribution in the diffraction pattern, which is the basis for our method.
The intense femtosecond FEL light pulses enable taking a snapshot of the particle in time, thus allowing tracking of the evolution of the particle in a pump-probe measurement scheme. This technique has been recently used to study light induced dynamics in a variety of systems, e.g., structural changes in rare gas clusters~\cite{Gorkhover2016,Fluckiger2016,Ferguson2016,Nishiyama2019}, surface melting of metal clusters \cite{Dold2020}, plasma dynamics in SiO$_2$ nanospheres \cite{Peltz2022}, and anisotropic evaporation of helium nanodroplets~\cite{Bacellar2022}. 
Here, we present a scenario where the energy is coupled into the system via the dopants at localized positions. 
The agreement between outstanding features in our recorded diffraction patterns and our qualitative model of randomly distributed and ordered voids in the droplets gives a strong indication that the dynamics is not homogeneous but starts at several distinct sites in the droplets.

\section{Methods}
\label{sec:methods}
The investigation is based on the analysis of wide-angle scattering images recorded at the FERMI FEL in a similar configuration to that previously described \cite{Langbehn2018}.
The experimental setup is described in section \ref{sec:exp_setup}. In section \ref{sec:scat_sim}, the procedure to simulate wide-angle diffraction patterns in order to retrieve information on the droplet dynamics from the recorded scattering images is presented.

\subsection{Pump-probe imaging setup}
\label{sec:exp_setup}
The experiment was carried out in a pump-probe measurement scheme at the low density matter (LDM) end-station \cite{Lyamayev2013} of the FERMI FEL. The helium nanodroplets are first irradiated with an NIR pulse (wavelength $\lambda_\mathrm{NIR} = 780~\mathrm{nm}$, duration $90~\mathrm{fs}$) to trigger dynamics (\emph{pump}) and subsequently, after a variable time delay of up to $800~\mathrm{ps}$, imaged using a $90~\mathrm{fs}$ long XUV pulse (\emph{probe}). The NIR laser is focused to a $100~\mathrm{\mu m}$ spot ($1/\rme^2$ diameter) resulting in a power density of $I_\mathrm{NIR}=9\times10^{13}~\mathrm{W}\mathrm{cm}^{-2}$.
The FEL is tuned to either non-resonant or resonant photon energies ($E_\mathrm{ph} = 19.4~\mathrm{eV}$ or $21.5~\mathrm{eV}$, respectively) delivering power densities exceeding $I_\mathrm{FEL}=3\times10^{14}~\mathrm{W}\mathrm{cm}^{-2}$ at spot sizes of $9\times13~\mathrm{\mu m}^2$ (FWHM) and smaller. 
Please note that the focus size of the NIR laser is chosen to be larger than that of the FEL to ensure that each droplet is hit by the NIR laser before being imaged with the FEL.
The XUV light scattered off a helium droplet is collected up to a maximum scattering angle $\theta_\mathrm{max}=30$\textdegree{} by a detector consisting of a circular microchannel plate (MCP) stacked onto a phosphor screen \cite{Bostedt2010} located $65~\mathrm{mm}$ downstream of the interaction region. Pictures of the amplified scattering images on the phosphor screen are taken via a 45\textdegree{}  mirror using a sCMOS Andor Neo 5.5 camera. The MCP, phosphor screen, and mirror have centered holes to let the FEL and NIR beams pass through.

The helium droplets are produced by supersonic expansion into vacuum of helium through a $100~\mathrm{\mu m}$ trumpet-shaped nozzle with half opening angle of 20\textdegree{} at a stagnation pressure $p_0=80~\mathrm{bar}$ and temperature $T_0=5.4~\mathrm{K}$, yielding a measured droplet velocity of $320~\mathrm{m}\mathrm{s}^{-1}$ \cite{Langbehn2018Supp} and a mean droplet radius of $\langle R \rangle = 400~\mathrm{nm}$ (corresponding to a mean droplet size of $\langle N \rangle = 6\times 10^9$ atoms per droplet) as determined from scattering images of approximately spherical droplets \cite{Langbehn2018Supp}. The droplets get doped with xenon by traversing a $35.3~\mathrm{mm}$ long gas cell, successively capturing $\sim1.6\times 10^6$ xenon atoms per droplet~(0.3~\textperthousand{})~\cite{Lewerenz1995,Toennies2004}. The kinetic energy deposited in the droplet by each captured atom leads to the evaporation of roughly 500 helium atoms \cite{Lewerenz1995,Fugol1978}. Hence, the average droplet radius can be calculated to shrink to about $\langle R' \rangle = 380~\mathrm{nm}$~\cite{Langbehn2021}, which matches well the measured average radius of doped droplets. Overall, the dopant atoms constitute only a small fraction of the total count of droplet atoms. Thus, for the wavelengths used in this experiment we can assume the recorded diffraction patterns are dominated by light scattered off the helium atoms.

In the analysis presented here, $26\,390$ images exhibiting meaningful scattering signal have been included.
The visibility of the scattering signal has been improved in post processing of the data by subtracting the straylight background. 
In addition, the images have been corrected for the uneven detector sensitivity (that is, e.g., due to the 8\textdegree{} bias angle of the MCP \cite{Fukuzawa2016}) and an angle-dependent intensity correction ($\propto \cos^{-3} \theta$) has been applied that accounts for the flat shape of the detection locus \cite{Bostedt2012}. 
The thus corrected patterns are made available via the CXI Data Bank \cite{Maia2012} under the identifier (ID) 208~\cite{CXIDB208}.

\subsection{Wide-angle scattering simulations}
\label{sec:scat_sim} 
In order to retrieve information on the droplet density distribution, the diffraction patterns of individual droplets are analyzed. In the case of \emph{small-angle} scattering ($\theta \lesssim 5$\textdegree{}) the droplet's electron density projection can be reconstructed from the diffraction pattern using a phase retrieval algorithm which is commonly utilized in CDI experiments and has already been successfully applied to doped helium nanodroplets \cite{Tanyag2015,Jones2016}. In the case of \emph{wide-angle} scattering, however, multiple projection planes contribute to the diffraction pattern rendering the reconstruction of a single projected density using algorithms based on inverse 2D Fourier transform impossible. For the same reason, wide-angle scattering images contain valuable information on the three-dimensional droplet shape and orientation, that can be retrieved by matching simulated diffraction patterns to the experimental data, as has been shown for faceted silver clusters \cite{Barke2015}. This \emph{forward-fitting} technique has further been employed to investigate the shapes of spinning helium nanodroplets~\cite{Bernando2017,Rupp2017,Langbehn2018} and is adapted here to address density fluctuations inside the droplets via systematic simulations. The choice of employing systematic simulations instead of directly deriving the shape information from the diffraction patterns was enforced by the firm constraints of the data set. These are (i)~the large number of strongly varying complex patterns, impossible to manually analyze in an appropriate manner to generate reliable statistics of the abundances, (ii)~the nonlinear detector response together with the wide-angle detection which prevent object reconstruction by iterative phase retrieval, and (iii)~the large range of complicated features from inhomogeneous densities in the evolved droplets that could not be captured by generalized model shapes, which are needed for direct forward fitting.

For calculating the wide-angle scattering patterns, a fast algorithm based on the multi slice Fourier transform (MSFT)~\cite{Colombo2022} approach was used.
While its approximations neglect multiple scattering events, e.g., backscattering of the propagating electromagnetic field, material properties can be approximated via effective optical parameters even for photon energies close to the resonance \cite{Langbehn2018Supp}. These parameters have been determined by matching scattering simulations to diffraction patterns of spherical helium nanodroplets and comparing the radial intensity distribution to simulations based on Mie theory \cite{Mie1908} for a variety of droplet sizes and photon energies ranging from $19.0~\mathrm{eV}$ to $23.7~\mathrm{eV}$ \cite{Langbehn2018Supp}.\\

\section{Results}
\label{sec:results}

In the following, the measured scattering images are described and discussed. 
First, the emergence of \emph{dynamic features} in the patterns is exemplified, \i.e., of features that are not observed in static (XUV only) data. In section~\ref{sec:temp_evo}, the temporal evolution of the patterns is investigated, while in section~\ref{sec:classification} different classes of patterns are identified based on their salient features.

\begin{figure}
   \centering
    \includegraphics{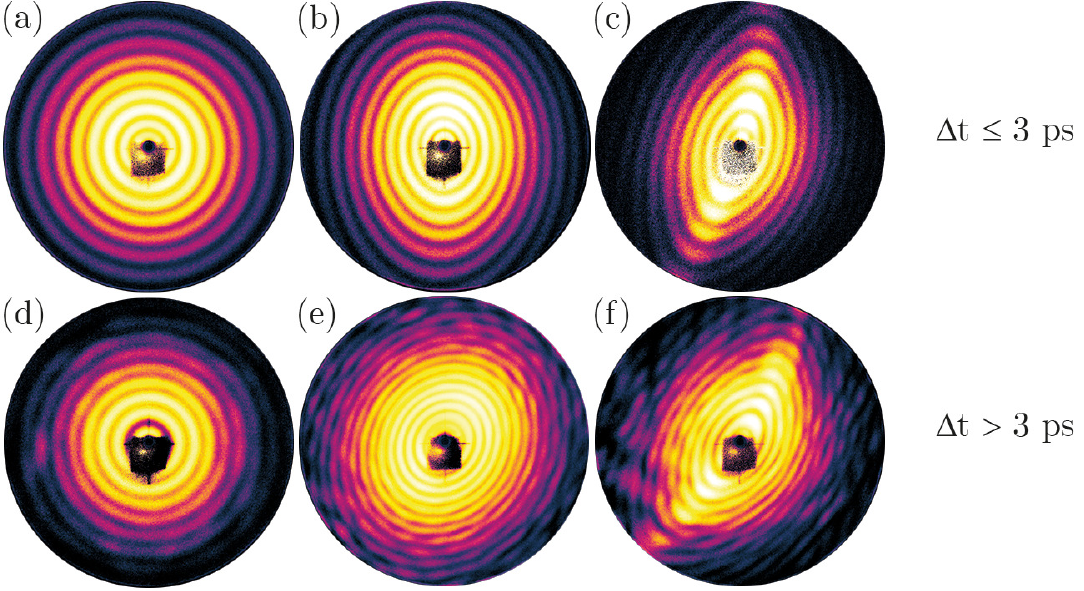}
    \caption{Emergence of dynamic features in diffraction patterns of xenon-doped helium nanodroplets. (a)-(c) For a time delay $\Delta t \leq 3~\mathrm{ps}$ the scattering images resemble those of pristine droplets described before \cite{Langbehn2018}, i.e., the dopants do not lead per se to visible changes in the patterns. (d)-(f) For longer delays, intensity fluctuations become apparent in the images that are linked to the dynamics in the droplets and therefore referred to as \emph{dynamic features}.}
    \label{fig:DynFeat}
\end{figure}

Figure~\ref{fig:DynFeat} shows a small selection of scattering images of xenon-doped helium nanodroplets at short time delay $\Delta t$ between pump and probe pulse. 
It should be noted that much more pronounced changes of the scattering occur at longer delays which will be discussed later (cf. figures~\ref{fig:BrightestHits} and~\ref{fig:DynChar}).
In the early stages of the dynamics, i.e., $\Delta t \leq 3\ \mathrm{ps}$, see figures~\ref{fig:DynFeat}(a)-(c), the images exhibit no visible difference from those of pristine droplets~\cite{Langbehn2018}: In contrast to data taken with hard x-rays \cite{Jones2016} the dopants do not lead to a change of the intensity distribution. Presumably, the nanometer-sized structures of dopant aggregates are not resolvable at the wavelength of the incoming XUV pulse.  
However, as shown in figures~\ref{fig:DynFeat}(d)-(f), intensity fluctuations along the diffraction rings become apparent for longer delays $\Delta t > 3\ \mathrm{ps}$. We assume these fluctuations are linked to the dynamics in the droplets and consequently refer to these data as images exhibiting \emph{dynamic features}. Details on the determination of the images showing dynamic features and an overview of the whole data set are given in~\ref{app:DynFrac}. Overall, for short time delays and only slight intensity fluctuations, the droplet shapes can still be inferred from the known characteristic patterns \cite{Langbehn2018} and classified as spherical [figures~\ref{fig:DynFeat}(a),(d)], ellipsoidal [figures~\ref{fig:DynFeat}(b),(e)], and pill-shaped [figures~\ref{fig:DynFeat}(c),(f)]. Nonetheless, the longer the delay the more complicated it is to identify droplet shapes, as the diffraction patterns get increasingly distorted (cf. figures~\ref{fig:BrightestHits} and~\ref{fig:DynChar}).

\subsection{Temporal evolution of the patterns}
\label{sec:temp_evo}

\begin{figure}
    \centering
    \includegraphics{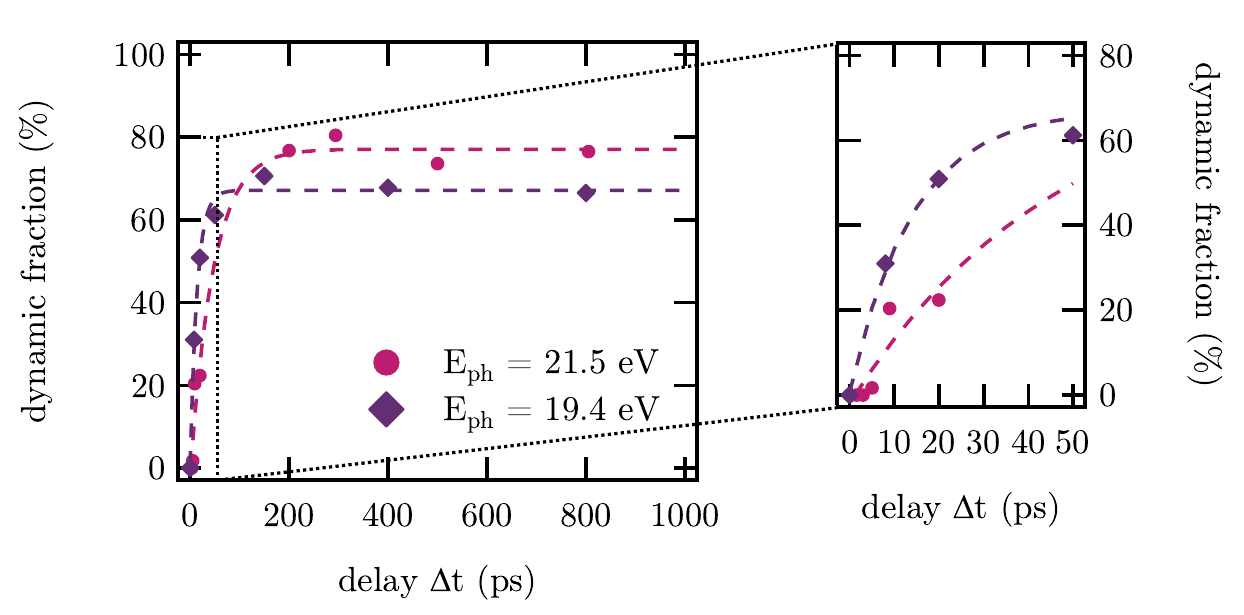}
    \caption{Temporal evolution of the fraction of scattering images exhibiting dynamic features. After a steep increase within $100~\mathrm{ps}$ to $150~\mathrm{ps}$ after NIR irradiation the dynamic fraction saturates at about 60\% to 80\%, i.e., the dynamics is not ignited in every droplet. From the inset it can be seen that for short delays ($\Delta t < 3~\mathrm{ps}$) no dynamics are observed in the patterns. Further, the rise of the dynamic fraction is slower for the resonant photon energy ($E_\mathrm{ph}=21.5~\mathrm{eV}$), cf. the data points at 20~ps. The dashed lines are limited growth functions as a guide to the eye.}
    \label{fig:DynFrac}
\end{figure}

In figure~\ref{fig:DynFrac}, the manually determined fraction of images exhibiting dynamic features is shown versus the time delay $\Delta t$ for the non-resonant ($E_\mathrm{ph}=19.4\ \mathrm{eV}$) and resonant ($E_\mathrm{ph}=21.5\ \mathrm{eV}$) photon energy of the XUV pulse. The reproducibility of the manually performed assignment of the dynamic fraction was tested (see~\ref{app:DynFrac}) to be better than 90\%. As a guide to the eye, limited growth functions are shown as dashed lines. 
Three main observations can be made: (i) There is a steep increase of the dynamic fraction, that is faster for the non-resonant wavelength, (ii) the dynamic fraction saturates to less than 100\% at long delays ($\Delta t > 200\ \mathrm{ps}$) indicating that not all images exhibit dynamic features, and (iii) the saturation percentage is higher for the resonant wavelength.

We attribute the first point to the wavelength-dependent influence of the refractive index on the scattering. In the resonant case ($E_\mathrm{ph}=21.5\ \mathrm{eV}$) the absorption is high, hence, mostly the front surface of the droplet contributes to the scattering and density changes inside the droplet are not visible in the diffraction pattern. When the dynamics initiated by the NIR pulse starts \emph{inside} the droplet (which we expect to be the case for xenon dopants) and propagates towards the surface, it can be observed earlier in the diffraction patterns recorded at the non-resonant photon energy ($E_\mathrm{ph}=19.4\ \mathrm{eV}$). The second and third observations, however, are most probably related to an experimental artifact: We assume that an imperfect overlap of the pulses leads to a situation where the dynamics is obviously not initiated in every droplet, even though a larger focal spot was chosen for the pump (NIR) than for the probe (XUV) pulse. Further, we identify the third observation -- a slightly lower asymptotic value for the non-resonant photon energy -- as a decreasing overlap between the two pulses over time, as even lower values were observed for subsequent delay scans that were taken several hours apart (not shown).

\begin{figure}
    \centering
    \includegraphics{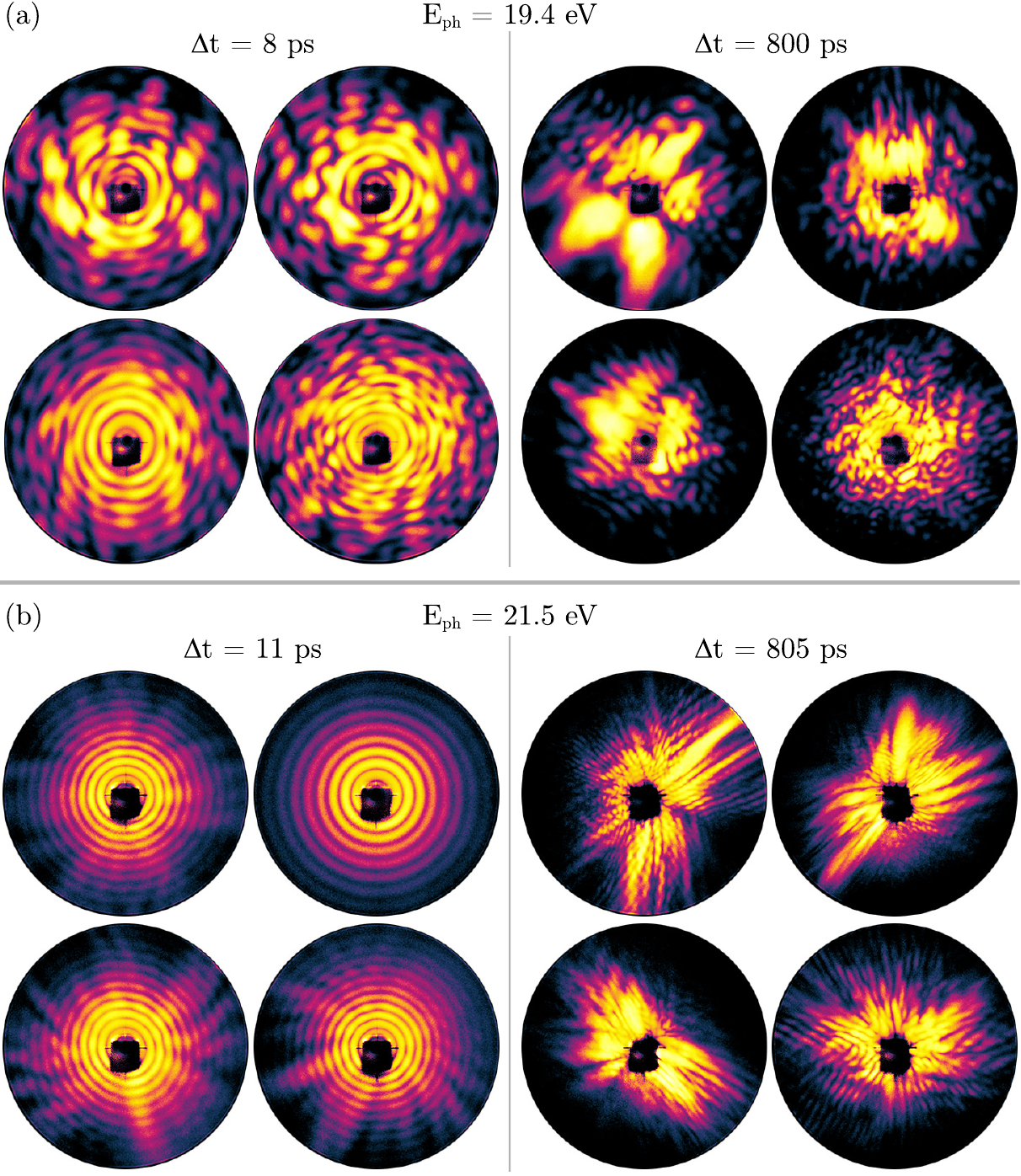}
    \caption{The brightest images of individual xenon-doped helium nanodroplets for short ($\sim$10~ps) and long ($\sim$800~ps) delays $\Delta t$ after irradiation with the NIR laser pulse. Although the excitation process is the same, the diffraction patterns look very different for the (a) non-resonant ($E_\mathrm{ph}=19.4~\mathrm{eV}$) and (b) resonant ($E_\mathrm{ph}=21.5~\mathrm{eV}$) photon energy. For example, the intensity fluctuations at short delays are much less pronounced for resonant scattering.}
    \label{fig:BrightestHits}
\end{figure}

The dynamic features in the patterns become, as expected, more pronounced with longer time delay. As well, the intensity not only fluctuates along the diffraction rings, but the complexity of the intensity distribution increases. It should be noted that many patterns exhibit strong directionality, which is completely different from what was observed in earlier pump-probe experiments on pure rare gas clusters \cite{Gorkhover2016,Fluckiger2016}, thus indicating multiple scattering centers at specific sites rather than surface smoothing or homogeneous fragmentation of the particle. The development towards long time scales is exemplified in figure~\ref{fig:BrightestHits}, where the four brightest scattering images are shown for short ($\sim$10~ps) and long ($\sim$800~ps) delays at non-resonant [figure~\ref{fig:BrightestHits}(a)] and resonant [figure~\ref{fig:BrightestHits}(b)] photon energy. It can be seen that the intensity fluctuations at short delays are more pronounced for non-resonant scattering ($E_\mathrm{ph}=19.4\ \mathrm{eV}$), where the droplets are almost transparent and information on density fluctuations inside the droplets is encoded in the diffraction pattern. For the resonant case ($E_\mathrm{ph}=21.5\ \mathrm{eV}$), less pronounced fluctuations presumably reflect changes on the droplet surface that occur later in time. Note that this observation is in line with the slower rise of the respective curve in figure~\ref{fig:DynFrac}.

The images shown in figure~\ref{fig:BrightestHits} already exhibit complex features, especially at long delays. However, this situation becomes even more complicated when not only the brightest images are examined but the whole data set is analyzed. 
While the overall trend -- the longer the delay, the more distorted the diffraction pattern -- remains the same, the degree of complexity in the images at a given delay varies considerably. In particular, no definite evolution in time is discernible when individual patterns are compared. Two facts can be considered to explain this observation: First, the scattering images reflect individual droplets of different size, orientation, and shape, leading to a huge variety of patterns. Second, as the position of the droplets with respect to the NIR and the FEL focus varies from shot to shot, both the dynamics triggered in the droplet and the intensity of the recorded scattering signal are affected. Finally, a decrease of the spatial overlap, e.g., because of a relative drift of the two beams, can further weaken the observable dynamics in the pattern. In order to address the complexity of the data set, we therefore further analyze the data by identifying classes of patterns with similar characteristic features and by discussing their origins.

\subsection{Pattern classification based on characteristic features}
\label{sec:classification}

\begin{figure}
    \centering
    \includegraphics{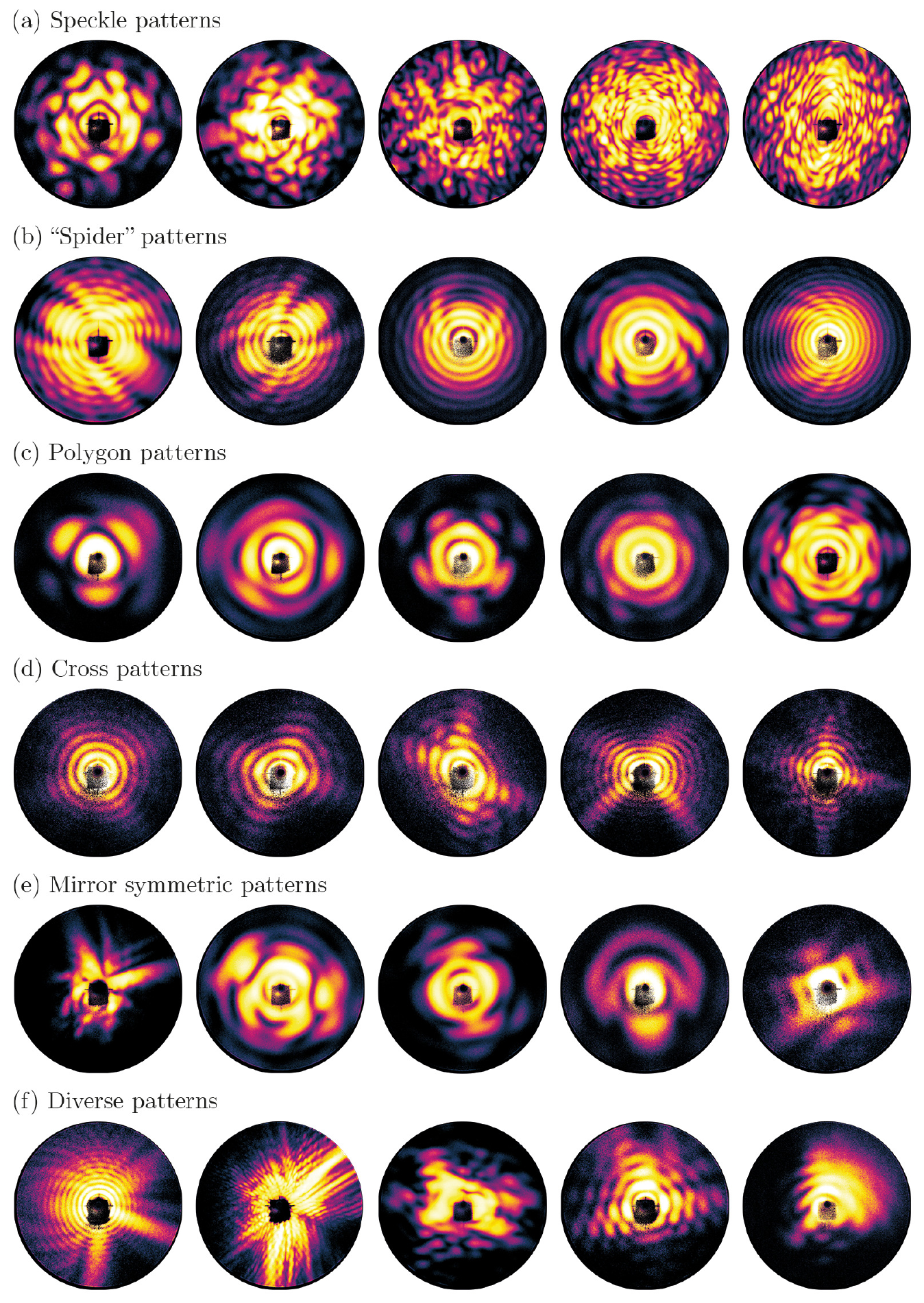}
    \caption{Classification of the dynamic data. Based on the characteristic features in the diffraction patterns the images are manually grouped into six categories: (a)~Speckle patterns, (b)~``spider'' patterns, (c)~polygon patterns, (d)~cross patterns, (e)~mirror symmetric patterns, and (f)~diverse patterns.}
    \label{fig:DynChar}
\end{figure}

In order to investigate the shapes of individual nanoparticles from diffraction patterns, it has previously proven helpful to group the scattering images into a few classes based on their characteristic features~\cite{Rupp2012,Rupp2014,Barke2015,Langbehn2018}. Here, we choose a similar approach to retrieve information about the dynamics in the droplets. This is based on the assumption that similar changes of the droplet density will lead to similar changes in the scattering images. We identify patterns whose characteristic features stand out on the data set and define classes for patterns exhibiting similar features. Figure~\ref{fig:DynChar} illustrates the variety as well as the complexity of the patterns in the data set. A selection of classes is shown from which the first three will be addressed in more detail in section~\ref{sec:model}. The first class, named \emph{speckle patterns}, comprises patterns exhibiting speckles of different sizes and is shown in figure~\ref{fig:DynChar}(a). The second class, shown in figure~\ref{fig:DynChar}(b), is called \emph{``spider'' patterns}, as streaks of higher intensity stick out of the pattern like pairs of ``legs'' on opposite sides or, in a more asymmetric version, on one side of the pattern. The third class consists of \emph{polygon patterns}, i.e., images exhibiting bright spots that define the corners of a polygon, see figure~\ref{fig:DynChar}(c). Two more classes are shown: Images exhibiting cross-like structures in figure~\ref{fig:DynChar}(d) and patterns with mirror symmetry, shown in figure~\ref{fig:DynChar}(e). Intriguingly, the latter induce the perception of shapes like, e.g., a butterfly, the face of a seal, or the greek letter $\Phi$, a phenomenon linked to human pattern recognition generally known as \emph{pareidolia} that might be in this case connected to the fractal characteristics of the patterns \cite{Taylor2017}. In addition, most patterns do not fit into one of the presented categories, examples are shown in figure~\ref{fig:DynChar}(f). It is important to note that despite this, the feature-based classification of the patterns makes it possible to develop an idea of the underlying structures that lead to similar characteristic scattering images, as we will exemplify in the following.

\section{Modeling}
\label{sec:model}

In order to deduce possible fluctuations in the droplets from the scattering images, we propose simple model shapes with fluctuating density that we use to simulate wide-angle diffraction patterns. The goal of the simulations is to qualitatively reproduce the features observed in the recorded images. To minimize the influence of absorption effects, we deliberately constrain our analysis to the non-resonant photon energy ($E_\mathrm{ph}=19.4\ \mathrm{eV}$).

As most of the helium nanodroplets exhibit spherical shapes (about 93\%, see \cite{Langbehn2018}), we choose a sphere as the basis of our model. We mimic fluctuations of the (optical) density in the droplet by introducing voids in the model shape in the form of small spheres that can be placed either randomly or at specific sites, e.g., along defined lines. This is motivated by the assumption that the dynamics in the droplets is triggered by the NIR laser field at the positions of the dopants, which presumably aggregate at multiple sites \cite{Loginov2011} or along vortex lines \cite{Gomez2014}. The diameter of the spherical voids is randomly changed by up to $\pm 20\%$ in order to avoid the occurrence of additional sharp features in the simulated diffraction pattern. It should be noted that the chosen representation of density fluctuations is not meant to imply any quantitative information on their optical properties that are, however, difficult if not impossible to determine. Nevertheless, also a qualitative comparison of the calculated intensity distribution and the characteristic features in the measured diffraction patterns enables tracing of structural changes occurring in the droplets.

In particular, we aim at a more detailed analysis of the structures leading to diffraction patterns in the speckle, ``spider'', and polygon classes [cf. figures~\ref{fig:DynChar}(a)-(c)]. 
While the intensity fluctuations in the speckle patterns [figure~\ref{fig:DynChar}(a)] seem to be randomly distributed, the ``spider'' and polygon patterns [figures~\ref{fig:DynChar}(b),(c)] exhibit pronounced directionality. We assume a connection between the fluctuations in the droplets and the intensity distribution in the diffraction patterns and therefore treat the former case, randomly distributed fluctuations, in section~\ref{sec:rand_fluct} and the latter case, structured fluctuations, in section~\ref{sec:struct_fluc}.

\subsection{Randomly distributed fluctuations (speckle patterns)}
\label{sec:rand_fluct}

\begin{figure}
    \centering
    \includegraphics{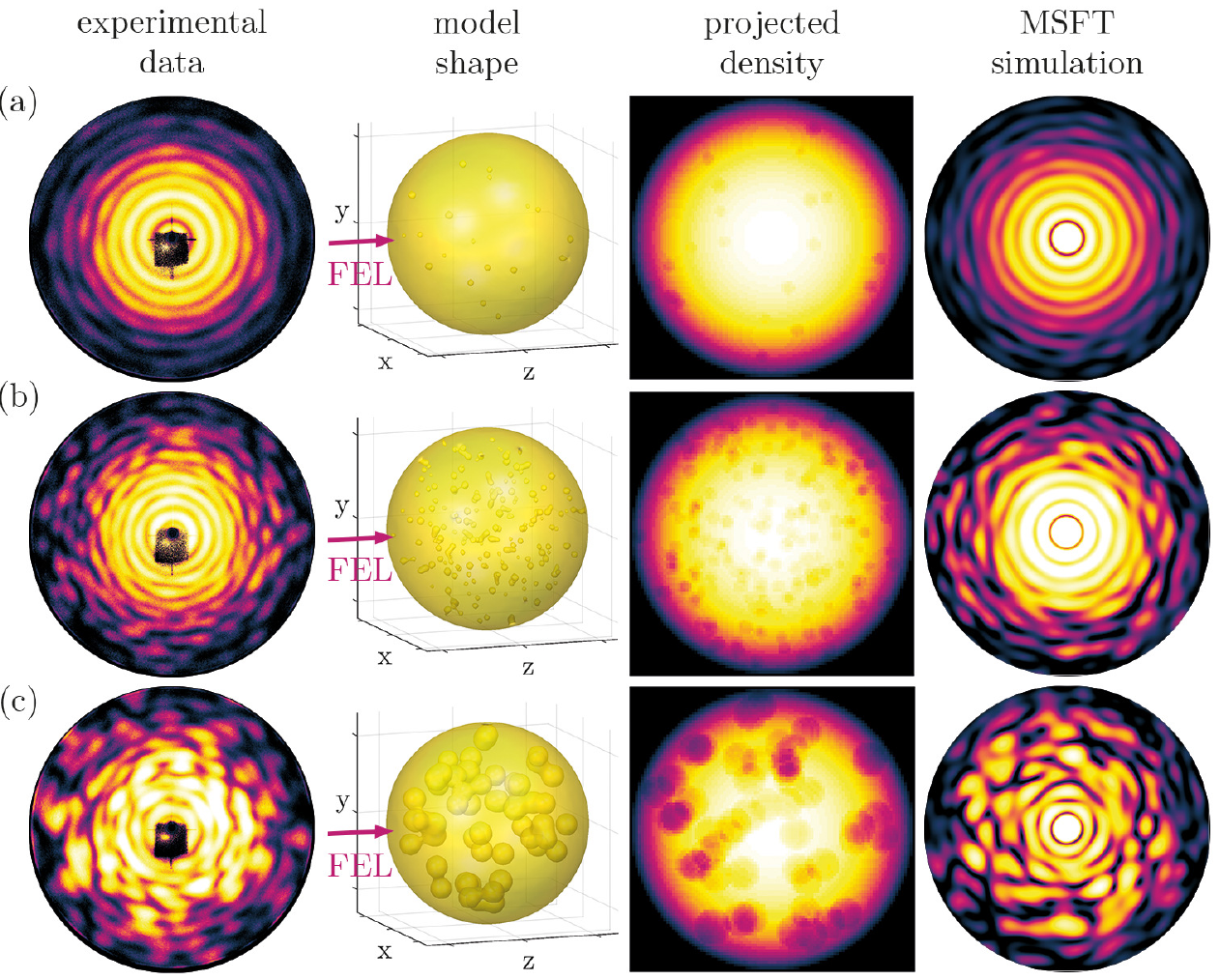}
    \caption{Speckle patterns indicating randomly distributed density fluctuations. In a simple model, we represent such fluctuations by voids or ``bubbles'' randomly distributed in the droplet. As can be seen from the 3D rendering of the model shape and its density projection, the number and size of the ``bubbles'' is varied from (a) to (c). The simulated diffraction patterns qualitatively reproduce the experimental data.}
    \label{fig:RandFluct}
\end{figure}

As has been discussed in section~\ref{sec:results}, intensity fluctuations in the scattering images start to occur at delays $\Delta t > 3\ \mathrm{ps}$ along the diffraction rings (cf. figure~\ref{fig:DynFeat}): From hardly visible fluctuations to slight variations of the intensity and a complete breakup of the rings, well-defined speckles emerge that are randomly distributed in the patterns. Figure~\ref{fig:RandFluct} shows experimental data of speckle patterns with increasingly pronounced fluctuations from (a) to (c).
Based on the idea of multicenter dopant aggregation in helium nanodroplets~\cite{Loginov2011}, we propose a sphere where bubble-like voids are introduced as a simple model representing random fluctuations in a droplet.
The number and size of the ``bubbles'' can be varied while their position in the droplet is chosen randomly. A 3D rendering of the model shape and its density projection along the FEL axis are also depicted in figure~\ref{fig:RandFluct}. Further, the corresponding simulated scattering image using the MSFT algorithm as described in section~\ref{sec:scat_sim} is shown. For a few and small ``bubbles'' [figure~\ref{fig:RandFluct}(a)], the diffraction rings are only slightly perturbed and the intensity fluctuations are more pronounced at larger scattering angles (i.e., towards the detector edge). When the number of ``bubbles'' is increased [figure~\ref{fig:RandFluct}(b)], the diffraction rings remain intact only at small scattering angles while towards the detector edge the rings get increasingly disrupted and speckles form. In the case of larger ``bubbles'' [figure~\ref{fig:RandFluct}(c)] the speckle size increases and the diffraction rings almost completely vanish, even for small scattering angles. It can be seen that the calculated diffraction patterns qualitatively reproduce the experimental data. This outcome indicates that the irregularities are randomly distributed in the droplet.

\subsection{Structured fluctuations (``spider'' and polygon patterns)}
\label{sec:struct_fluc}

\begin{figure}
    \centering
    \includegraphics{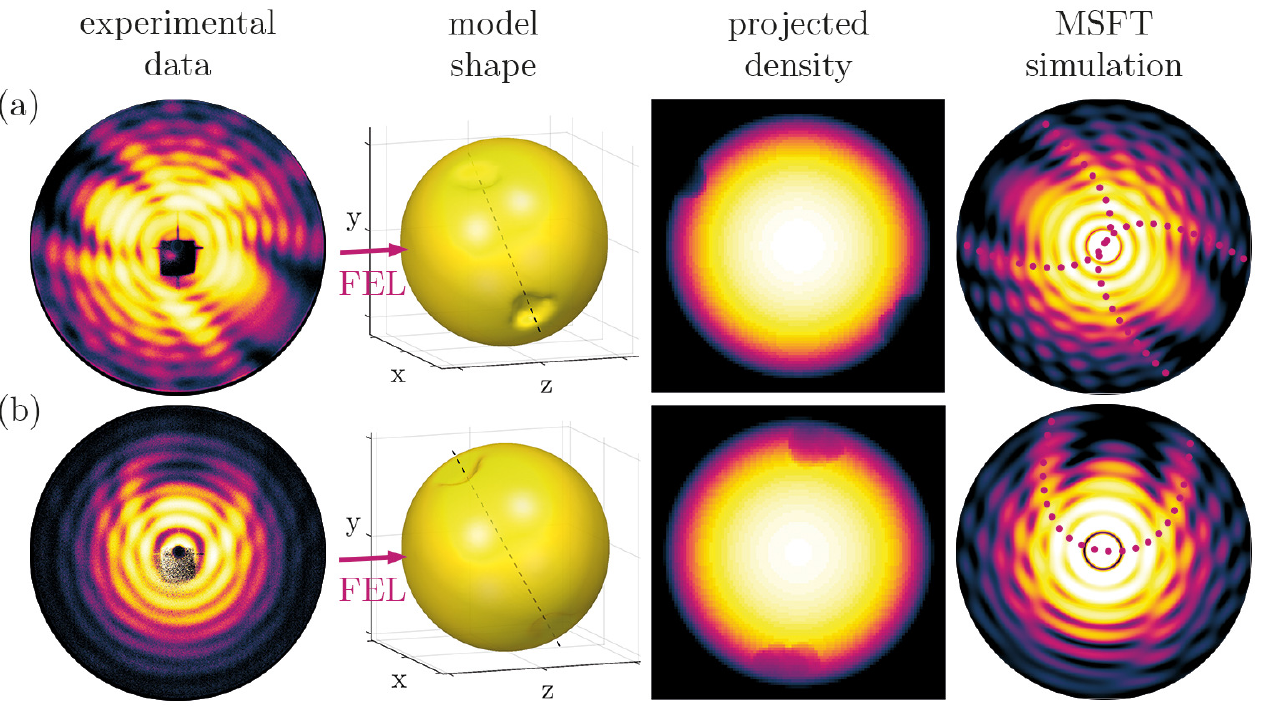}
    \caption{``Spider'' patterns indicating density fluctuations at specific sites on the droplet surface. We model such fluctuations by placing multiple, overlapping bubbles close to the droplet surface thus forming two dimples on opposite sides. (a) When the axis connecting the dimples (dashed line) is approximately perpendicular to the FEL axis (pink arrow), both dimples contribute to the scattering leading to spider-like legs on two sides of the pattern (dotted lines). (b) When the axis connecting the dimples is considerably tilted, mostly the dimple closer to the light source contributes to the scattering. Consequently, the spider-like legs are only visible to one side of the diffraction pattern (dotted line).}
    \label{fig:StructFluct}
\end{figure}

In the case of scattering images exhibiting intensity fluctuations with strong directionality, the diffraction patterns point to some underlying structure governing the density fluctuations in the droplets. For example, in superfluid helium nanodroplets, quantum vortices can dictate the arrangement of dopant structures~\cite{Gomez2014}. In the following, a model is presented that is able to reproduce the observed scattering images of the ``spider'' and the polygon classes [cf. figures~\ref{fig:DynChar}(b),(c)].

Two examples of ``spider'' patterns are shown in figure~\ref{fig:StructFluct}. The peculiar feature of these images is that they exhibit several pairs of streaks (reminiscent of spider-like ``legs'') with increased intensity towards larger scattering angles. In addition to the experimental data the proposed model shape and its density projection as well as the corresponding MSFT simulation are shown. The model is constructed by placing several overlapping bubbles on or close to the surface of a sphere, which leads to a spherical shape with dimples on opposite sides. When the axis connecting these dimples is approximately perpendicular to the FEL axis, as is shown in figure~\ref{fig:StructFluct}(a), the resulting diffraction pattern exhibits the characteristic, spider-like legs (see the dotted lines in the simulation). Figure~\ref{fig:StructFluct}(b) illustrates the case when the axis connecting the dimples is considerably tilted with respect to the FEL axis. In consequence, mostly the dimple closer to the light source contributes to the scattering, which we attribute to absorption effects along the beam path, and the spider-like legs occur only to one side of the pattern. The simulations qualitatively reproduce the experimental data. Hence, these scattering images might be caused by a single or a few irregularities at or close to the droplet surface.


\begin{figure}
    \centering
    \includegraphics{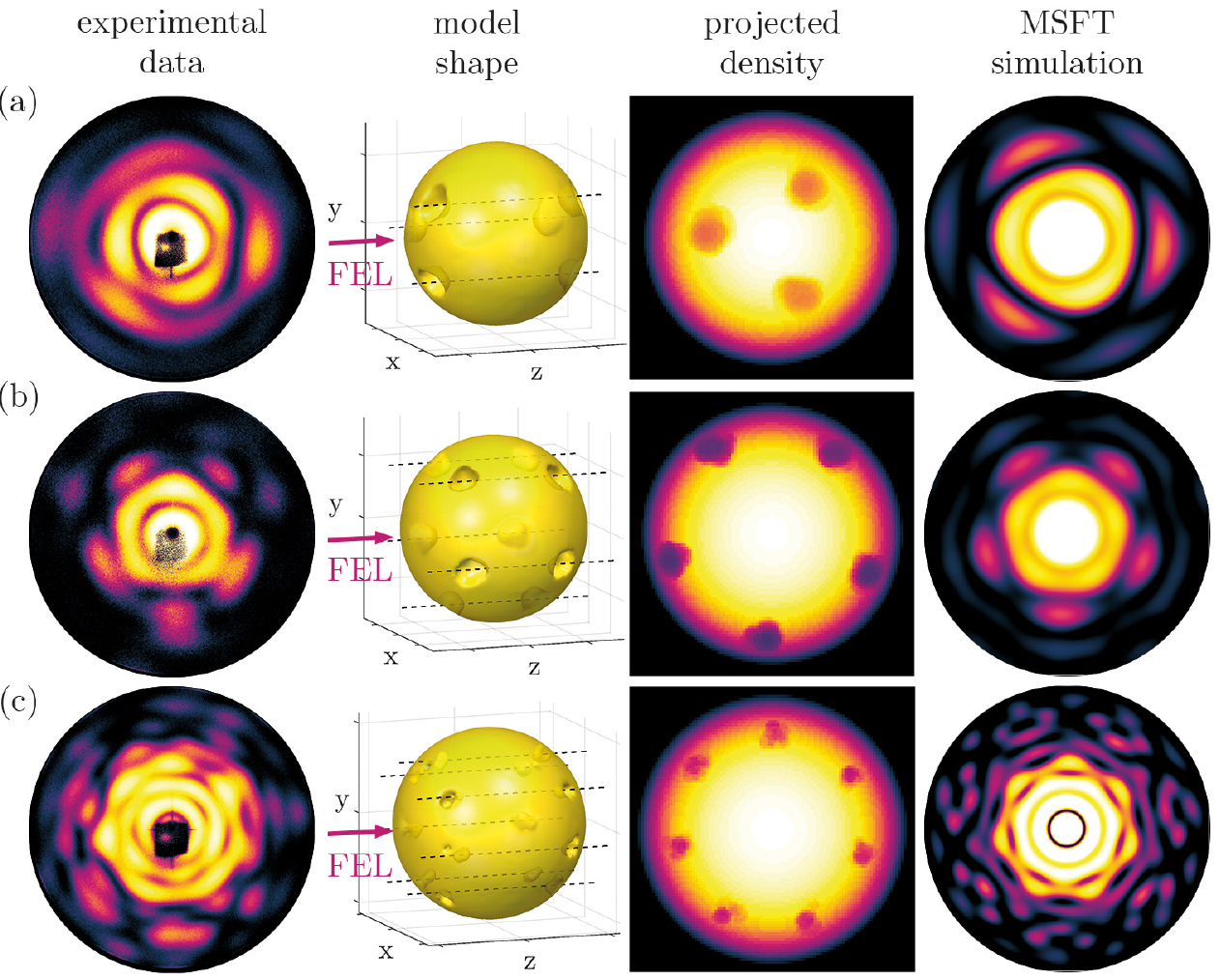}
    \caption{Polygon patterns indicating a regular arrangement of the density fluctuations. The patterns can be reproduced when the dimples are placed in a polygon configuration while the axes connecting the dimples (dashed lines) are oriented parallel to the FEL axis.  
    The calculated diffraction patterns for (a) triangular, (b) pentagonal, and (c) heptagonal configurations qualitatively reproduce the experimental data.}
    \label{fig:VortexFluct}
\end{figure}

Another conspicuous image type is that of polygon patterns, where bright spots can be seen that are arranged like the corners of a polygon. Figure~\ref{fig:VortexFluct} shows examples for (a)~triangular, (b)~pentagonal, and (c)~heptagonal configurations. We use the same model as before but with multiple dimples on the droplet surface in a polygonal arrangement. The experimental data are qualitatively reproduced by the simulations when the axes connecting the dimples are approximately parallel to the FEL axis. While the positions of the dimples are difficult to identify in the 3D model representation, they are clearly visible in the density projection. Overall, dimples at specific sites on the droplet surface can explain the observed diffraction patterns. While the ``spider'' patterns point to two dimples on opposite sides of the droplet, the polygon patterns imply a regular arrangement of multiple dimples. A possible origin for these structures can be a grid-like distribution of the dopants in the droplets, as will be discussed in the following section.

\section{Discussion}
\label{sec:discussion}

Our observations of dynamics in xenon-doped helium nanodroplets are very different from what has been previously seen for pure clusters. In contrast to pristine systems, where nanoplasma dynamics linked to the NIR laser field have been reported to lead to surface melting \cite{Gorkhover2016} or anisotropic evaporation \cite{Bacellar2022}, the diffraction patterns of the heteronuclear system investigated here exhibit complex features that point to fluctuations in the droplets, initiated at the dopant positions. A huge variety of patterns is observed making it difficult to assign their features to possible droplet configurations and their specific temporal evolution. Nevertheless, we identified characteristic patterns that suggest a random or a structured distribution of the dopants in the droplets. These observations lead to three directions we want to further explore in the following discussion: (i)~What is the underlying process of the fluctuations? (ii)~Do the fluctuations occur in the volume or at the surface of the droplets? (iii)~How do the fluctuations evolve over time?

Up to now, theoretical studies of the interaction of intense NIR pulses with doped helium nanodroplets have addressed the nanoplasma dynamics that evolve in the droplets within a few hundreds of femtoseconds after irradiation \cite{Mikaberidze2008,Mikaberidze2009,Peltz2011,Krishnan2011,Krishnan2012}. The dynamics we observed in the diffraction patterns are on a much longer timescale, which might be a direct result of the lower NIR power density in our experiment (although the parameters -- fraction of dopants in the droplet and power density -- are comparable to those given in \cite{Mikaberidze2009}).
The scattering images remain unchanged for at least 3 ps, which is already a long time compared to the calculated nanoplasma dynamics, and images exhibiting bright scattering signal could be recorded up to 800 ps. On this timescale we would expect that to a large extent recombination processes have already taken place while the particle has fragmented~\cite{Fluckiger2016}. Therefore, it is questionable whether we directly observe the forming nanoplasma in the droplets, i.e., fluctuations of electronic origin. Another possibility is that fluctuations of the droplet density occur in the form of a geometric displacement of the droplet atoms, e.g., because of the formation of gas bubbles~\cite{Fernandez2015,Jones2019}. In the case of multiple dopant clusters randomly distributed in the droplet, this could lead to the growth of multiple gas bubbles that become visible in the diffraction patterns as speckles after a certain time, when they are large enough to be detected.
Note that the observed speckle patterns could also be caused by strong fluctuations on the droplet surface. However, xenon dopants are immersed in the droplets and do not remain on the surface~\cite{Toennies2004}; further, surface irregularities large enough to cause distortions of the diffraction rings have not been observed for pure droplets~\cite{Bacellar2022}. Therefore, we assume the fluctuations take place inside the droplets. In addition, the fact that the dynamics in this experiment is observed earlier at the non-resonant photon energy, where more information from the droplet interior contributes to the scattering, could be a hint that fluctuations indeed start inside the droplet, not on the surface. On the other hand, the ``spider'' patterns clearly indicate that at least in some cases the fluctuations happen at or close to the surface. While this could be in principle caused by bubbles propagating towards the surface (that could eventually get ejected from the droplet), there is no clear temporal evolution from patterns reflecting irregularities inside the droplet to patterns showing changes on the droplet surface, i.e., speckle and ``spider'' patterns occur at the same time delay. 

We therefore discuss another mechanism that has been described before: The ejection of dopants from a helium droplet when they are ionized~\cite{Bonhommeau2008,Halberstadt2020}.
In this case, dopant atoms that aggregated close to the droplet surface probably get ejected earlier and some helium atoms might be dragged away by the dopants, thus leading to the formation of dimples on the surface. Furthermore, dopants arranged along an array of vortices might therefore lead to a regular arrangement of such fluctuations. In particular, the strong resemblance of the polygon patterns observed in this experiment to previously reported diffraction patterns of superfluid helium nanodroplets hosting quantized vortices \cite{Jones2016} suggests a connection between the vortex array and the density fluctuations. 

Next, we like to make some comments on the challenges of our analysis and how to overcome them. 
The three guiding questions of our discussion, (i)~to identify an underlying process, (ii)~to determine the locations of the fluctuations, and (iii)~to develop a picture of the evolving structures over time, can only be partially answered by the analysis presented here: Certainly, in order to substantiate our initial ideas on the dynamic processes in the droplets and ultimately answer the questions raised by our analysis, theoretical studies addressing both nanoplasma dynamics and structural fluctuations are needed, especially for large droplets hosting multiple dopant clusters or dopant structures that formed along vortex lines. Our results underline that such calculations should span a time scale of several hundred picoseconds. 
The temporal evolution of the fluctuations is even more difficult to analyze due to the large variety of diffraction patterns, the lack of an obvious time ordering of the observed patterns, and the extensive size of the data set. 
Our analysis is focused on a small subset of distinctive patterns to develop first ideas on the processes following irradiation with intense light pulses. In order to trace the evolution of the dynamics, however, one needs to identify patterns that are characteristic for a specific droplet configuration at each delay. This is a very challenging and time consuming task that might not even be possible to perform manually. Instead, it could be fruitful to analyze the whole data set by using an automated approach for pattern recognition, e.g., a neural network, as has been previously used for classification of static diffraction patterns \cite{Zimmermann2019}. In particular, developing an unsupervised network that can identify classes of characteristic patterns for specific delays as an unbiased assistant to the human researcher would be an interesting prospect for future work. 
In combination with an automated forward-fitting routine enabling a quantitative 3D shape retrieval, e.g., based on a set of orthogonal functions and an efficient and fast MSFT algorithm~\cite{Colombo2022}, the temporal evolution of the droplet shapes after NIR irradiation could be finally revealed.

\section{Conclusion}
\label{sec:summary}

To summarize, we observed the dynamics in xenon-doped helium nanodroplets after irradiation with an intense NIR pulse by analyzing wide-angle scattering images recorded in a pump-probe experiment at the FERMI FEL. Our data show the dynamics is linked to the xenon structures that formed in the droplets and further corroborate the idea that the droplets do not expand homogeneously. 
As the time scale of the observed dynamics is remarkably long (up to 800 ps), we believe its origin to be rather geometric than electronic.

We analyzed the data by identifying scattering images that show distinctive intensity distributions and grouping them into different classes based on characteristic features in the diffraction patterns. In particular, we associate scattering images exhibiting speckle patterns or ``spider'' and polygon patterns to fluctuations that are randomly distributed in the droplet or that occur at specific sites on or close to the droplet surface. We propose that the fluctuations are connected to either multiple dopant clusters randomly distributed in the droplet or dopants that agglomerate along quantized vortices in array-like structures. In order to distinguish different droplet configurations and trace their temporal evolution, it will be very promising to establish a 3D-sensitive shape retrieval algorithm and an automated classification of the scattering images, which are under development.

Our results suggest that in addition to studies of nanoplasma evolution, processes describing structural fluctuations such as bubble formation~\cite{Fernandez2015,Jones2019} or dopant ejection~\cite{Bonhommeau2008,Halberstadt2020} should be further investigated both experimentally and theoretically, especially for the case of large doped helium nanodroplets. 
As the considerable amount of energy deposited in the droplets by the NIR pulse does not lead to their immediate destruction, additional dissipation channels associated to the superfluid nature of the droplets should be included in the models.
In this context, a better understanding of the underlying processes contributes to the prospect of using helium nanodroplets as sacrificial layers in CDI experiments~\cite{Hau-Riege2007,Mikaberidze2008,Chapman2009} or as an ultracold environment for controlled nanostructure growth. Finally, using an NIR laser pulse to magnify the structure of an embedded particle by creating gas bubbles at the positions of its atoms, thus making it possible to analyze the configuration of nanometer-sized objects even at relatively long (i.e., XUV) wavelengths, is an interesting approach, especially in the context of laboratory based experiments using high harmonic generation sources for single particle imaging~\cite{Rupp2017}.

\ack
We would like to acknowledge excellent support by the FERMI staff during the beam time. This work received financial support from the Deutsche Forschungsgemeinschaft under the Grant No. MO 719/14-1, from the Leibniz-Gemeinschaft under the Grant No. SAW/2017/MBI4, and from the Swiss National Science Foundation under the Grant No. 200021E\_193642 as well as via the NCCR MUST program. F.S. acknowledges funding by Deutsche Forschungsgemeinschaft under the Grant No. STI 125/19-2.

\section*{Data availability statement}
The complete data set is available at the Coherent X-Ray Imaging Data Bank (CXIDB)~\cite{Maia2012} under the identifier (ID) 208: \url{https://cxidb.org/id-208.html}

\appendix
\section{Determination of the dynamic fraction}
\setcounter{section}{1}
\label{app:DynFrac}

\begin{figure}
   \centering
    \includegraphics{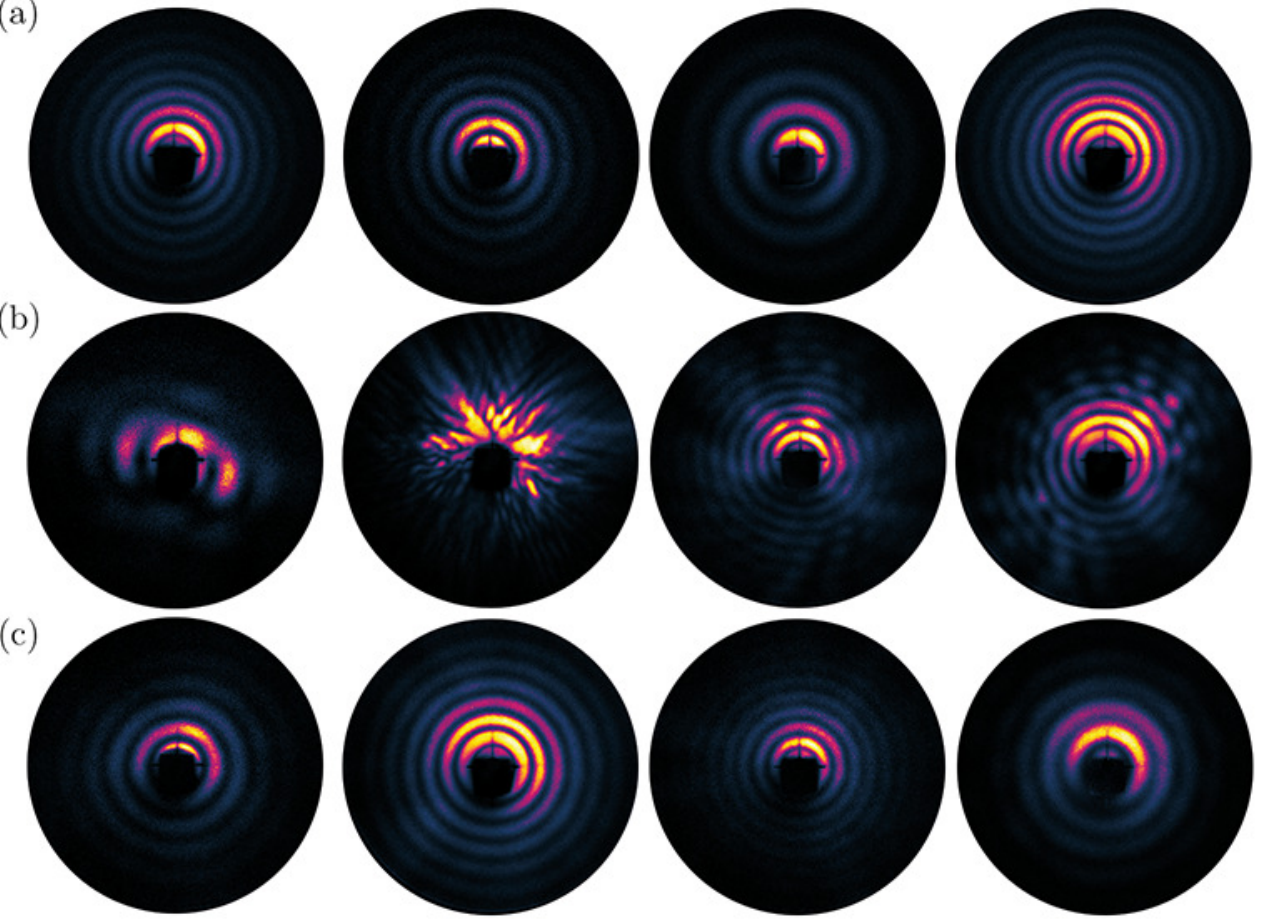}
    \caption{Classification of static / dynamic images. (a) Scattering images with no distortion of the diffraction rings are labelled ``static''. (b) Images exhibiting changes in the diffraction are labelled ``dynamic''. (c) Examples for images showing only subtle changes in the diffraction rings. Note that for the images shown here, the correction for the uneven detector sensitivity (cf. section~\ref{sec:exp_setup}) has not been applied.}
    \label{fig:DynClass}
\end{figure}

In the following, we exemplify the classification of the data set to determine the dynamic fraction (cf. figure~\ref{fig:DynFeat}), which is a measure for the number of images that exhibit changes in the diffraction pattern that are due to the prior irradiation with a strong NIR pulse. An overview of the data used for this analysis is given in table~\ref{tab:DataOverview}. For each measurement run, the images showing significant scattering signal (i.e., the ``hits'') are filtered out by simply thresholding the integrated intensity of the image. In a second step, out of the total $26\,390$ hits, the images exhibiting dynamic features are manually identified by the researcher. These are images that show features that are not observed for the scattering images with only the XUV pulse present (i.e., ``static'' data). This is further exemplified in figure~\ref{fig:DynClass}. The images shown in the first row, figure~\ref{fig:DynClass}(a), show very smooth diffraction rings that are not distorted, as it is known for static diffraction imaging data of superfluid helium nanodroplets (cf. the scattering images presented in~\cite{Langbehn2018}). In contrast, figure~\ref{fig:DynClass}(b) shows examples of images exhibiting some kind of distortion of the rings, hence, these images are labeled ``dynamic''. In order to illustrate the confidence of the classification, figure~\ref{fig:DynClass}(c) shows images exhibiting very subtle changes in the diffraction. They cannot be easily assigned to one of the two classes, as it remains unclear whether the distortion of the rings is caused by dynamic effects in the droplets or by, e.g., inhomogeneities of the detector sensitivity. To give an estimate of the accuracy of our classification, we additionally labeled a set of $1\,000$ randomly chosen scattering images as ``static'' or ``dynamic'' and compared the results to the initial classification. For 93.6\% of the images, the labels agree with the initial assignment, while 4.8\% of the images were initially labeled wrong as ``static'' although they show clear dynamic features. For 1.6\% of the images the changes in the patterns are too subtle to decide whether they show dynamic features or not.

\begin{table}
\caption{Data overview. For each measurement run, the images showing significant scattering signal (i.e., the ``hits'') are identified. The images exhibiting dynamic features are manually determined by the researcher. For details, see text.}
\label{tab:DataOverview}
\footnotesize
\begin{tabular}{@{}llllll}
\br
Run No. & $E_\mathrm{ph}$ (eV) & $\Delta t$ (ps) &  No. of shots & No. of hits & No. of images exhibiting\\ 
 & & & & & dynamic features\\
\mr
153 & 21.5 & 200 & 12\,000 & 1\,897 & 1\,455\\
155/156 & 21.5 & 500 & 12\,000 & 1\,704 & 1\,254\\
157 & 21.5 & 805 & 12\,000 & 1\,555 & 1\,189\\
162 & 21.5 & 294.3 & 11\,300 & 1\,391 & 1\,120\\
163 & 21.5 & 0 & 12\,000 & 1\,458 & 0\\
164 & 21.5 & 9 & 12\,000 & 1\,601 & 327\\
165 & 21.5 & 1.5 & 12\,000 & 1\,677 & 0\\
166 & 21.5 & 5 & 12\,000 & 1\,636 & 27\\
167/168 & 21.5 & 3 & 12\,000 & 1\,602 & 0\\
169/170 & 21.5 & 20 & 12\,000 & 1\,676 & 376\\
175 & 19.4 & 0 & 10\,000 & 1\,429 & 0\\
176 & 19.4 & 8 & 10\,000 & 1\,773 & 550\\
177 & 19.4 & 50 & 10\,000 & 1\,667 & 1\,020\\
178 & 19.4 & 800 & 10\,000 & 1\,105 & 735\\
179 & 19.4 & 20 & 10\,000 & 1\,597 & 813\\
180 & 19.4 & 150 & 10\,000 & 1\,407 & 993\\
181 & 19.4 & 400 & 10\,000 & 1\,215 & 824\\
\br
\end{tabular}
\end{table}
\normalsize

\section*{References}
\printbibliography[heading=none]

\end{document}